\documentclass[a4paper]{VisionStyle}
\usepackage{epsfig}

\begin{document}

\title{Helium-like triplet diagnostics}

\author{J. Dubau\inst{1,3}  \and D. Porquet\inst{2,3} } 

\institute{
LSAI, U.M.R. 8624, CNRS, Universit{\'e} de Paris Sud, 91405 Orsay Cedex, France
\and 
Service d'Astrophysique, CEA Saclay, 91191 Gif-sur-Yvette Cedex, France
\and 
LUTH, F.R.E. 2462 CNRS, Observatoire de Paris, 92195 Meudon Cedex, France
}

\maketitle 

\begin{abstract}
 The 1s$^{2}$--1s2l  lines are the most intense He-like ions lines. 
They are used as spectroscopic diagnostics for solar active regions 
as well as for different laboratory plasmas. 
Nowadays, it exits very 
high spectral resolution instruments and, for intense X-ray sources, 
one  can do spectroscopic diagnostics from line ratios. With XMM (RGS) 
and Chandra (LETGS, HETGS) spectral resolutions and for several atomic elements, 
it is particularly possible to separate a 3 blended line set, the so-called He-like triplet: 
Resonance (r), Intercombination (i) and Forbidden (f), which are dominated 
respectively by lines issued from the following levels : 
1s2p $^{1}$P$_{1}$, 1s2p $^{3}$P$_{1,2}$, and 1s2s$^{3}$S$_{1}$. 
We shall show that the measurement of two different ratios between 
these 3 lines (R = f/i and G = (f + i)/r) give quantitative informations 
on the nature of the emitting plasma (photo-ionized or collisional) 
and on its electronic density and temperature.
A more refined analysis must also include satellite line contributions. 

\end{abstract}

\section{Introduction}

\begin{figure}[!t]
\epsfig{file=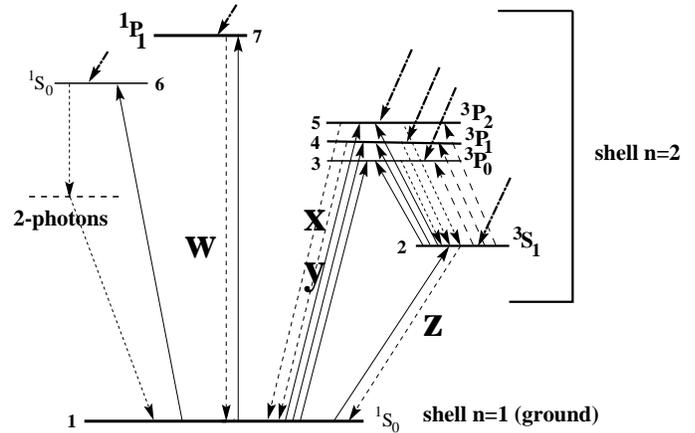,width=8.9cm}
\caption{Simplified level scheme for Helium-like ions. {\bf w} (or r), 
{\bf x,y} (or i), and {\bf z} (or f): resonance, intercombination, and 
forbidden lines, respectively. {\it Full upward arrows}: 
collisional excitation transitions, {\it broken arrows}: 
radiative transitions (including photo-excitation 
from 2\,$^{3}$S$_{1}$ to 2\,$^{3}$P$_{0,1,2}$ levels, 
and 2-photon continuum from  2\,$^{1}$S$_{0}$ to the ground level), 
and  {\it thick skew arrows}:  
recombination (radiative and dielectronic) plus cascade processes.}
\label{fig:gotrian}
\end{figure}

\indent In the X-ray range, the three most intense 
lines of Helium-like ions (``triplet'') are: 
the {\it resonance} line ($r$, also called $w$: 
1s$^{2}$\,$^{1}S_{\mathrm{0}}$ -- 1s2p\,$^{1}P_{\mathrm{1}}$), 
the unresolved {\it intercombination} lines 
($i$, also called $x+y$: 1s$^{2}$\,$^{1}S_{0}$ -- 1s2p\,$^{3}P_{2,1}$) and 
the {\it forbidden} line ($f$, also called $z$: 1s$^{2}$\,$^{1}{S}_{0}$ -- 1s2s\,$^{3}{S}_{1}$). 
They correspond to transitions between the $n$=2 shell and the $n$=1 ground-state shell 
(see Figure~\ref{fig:gotrian}).\\ 

Gabriel \& Jordan (\cite{jdubau-WB2:Gabriel69}) were the first 
to recognize that the ratios of these lines  could give important 
plasma diagnostics on the electron density ($n_{e}$) and temperature ($T_{e}$). 
\begin{equation}\label{eq:RG}
R~(n_e)~=~\frac{f}{i}~~~~~~~~~~~~~~G~(T_e)=\frac{f + i}{r}
\end{equation}

These diagnostics have been used for coronal spectra 
as well as for many laboratory hot plasmas (e.g. tokamaks). \\

Helium-like spectral lines 1s$^{2}$-1s2l have been observed 
with high spectral resolution for a long time in 
solar corona X-ray spectra. 
During the seventies, several instruments have been flown aboard
satellites to observe the soft X-ray spectra of
solar flares. Spectra in the 1--25\AA~ range, for example, have been obtained 
from uncollimated crystal spectrometers on the early {\sl OSO} satellites 
(Neupert et al. \cite{jdubau-WB2:Neupert73}, 
Doschek et al. \cite{jdubau-WB2:Doschek72}), the {\sl OVI-17} 
satellite (Walker \& Rugge \cite{jdubau-WB2:Walker70}), and 
{\sl OSO 8} (Parkinson et al. \cite{jdubau-WB2:Parkinson78}). Particular spectral regions
(e.g. iron lines at 1.9\AA) have been examined with the 
{\sl intercosmos-4} spectrometer (Grineva et al. \cite{jdubau-WB2:Grineva73}) 
and the {\sl SOLFLEX} instrument on the {\it P78-1} 
satellite (Doschek et al. \cite{jdubau-WB2:Doschek79}). 
Rocket-borne collimated crystal spectrometers
have been recorded spectra in various ranges up to
25\AA~ from active regions or weak flares 
(Parkinson \cite{jdubau-WB2:Parkinson75}; Pye et al. \cite{jdubau-WB2:Pye77}, 
Burek et al. \cite{jdubau-WB2:Burek81}), while the collimated {\sl SOLFLEX} instrument 
on {\sl P78-1} has been obtained very high resolution 
spectra from sizable flares (McKenzie et al. 1980).
Thanks to the Flat Crystal Spectrometer 
(FCS, Phillips et al. \cite{jdubau-WB2:Phillips82}) and Bent Crystal 
spectrometers (BCS, Culhane et al. \cite{jdubau-WB2:Culhane81})
on board  the {\sl Solar Maximum Mission (SMM)}, 
many line blends were resolved for the first time 
allowing new physical information about the emitting plasma to be extracted.
In particular the BCS observed He-like iron and calcium.
 More recently Yohkoh (Solar A) contained also a BCS which pursued 
similar solar observations of iron and calcium 
as {\sl SMM}, 
 extending the observations to lower temperature by
using Helium-like sulfur spectra 
(Harra-Murnion et al. \cite{jdubau-WB2:Harra-Murnion96}, 
Kato et al. \cite{jdubau-WB2:Kato97}).     

For active coronae and solar flares, 
 a temporal analysis of these lines can be carried on during 
the three consecutive phases: ionization, gradual, recombination 
(e.g., Mewe \& Schrijver \cite{jdubau-WB2:Mewe78a}, 
\cite{jdubau-WB2:Mewe78b}, \cite{jdubau-WB2:Mewe78c}; 
Doyle \cite{jdubau-WB2:Doyle80}; 
Pradhan \& Shull \cite{jdubau-WB2:Pradhan81}).\\

It is now possible, thanks to the spectral resolution of the
 new generation of X-ray satellites: Chandra and XMM-Newton, 
to resolve this triplet and to use these diagnostics in case 
of extra-solar objects.
 Indeed, the Helium-like ``triplet'' is a powerful tool 
in the analysis of high-resolution spectra of a variety of plasmas such as: \\
$\bullet$ collisional plasmas: 
{\it e.g., stellar coronae (OB stars, late type stars, active stars, ...)}\\
$\bullet$ photo-ionized or hybrid plasmas (photo-ionization + collisional ionization): 
{\it e.g., ``Warm Absorber'' (in AGNs), X-ray binaries, ...}\\
$\bullet$ out of equilibrium plasmas: {\it e.g., SNRs, stellar flares, ...}.\\

  The wavelength ranges of the RGS (6-35 {\AA}), of the LETGS (2-175 {\AA}), and 
of the HETGS (MEG range: 2.5-31 {\AA}; HEG range: 1.2-15 {\AA}) contain the Helium-like 
line "triplets'' from \ion{C}{v} (or \ion{N}{vi} for the RGS, and for the HETGS-HEG) 
to \ion{Si}{xiii} (Table~\ref{table:lambda}).

\begin{table}[!h]
\label{table:lambda}
\caption{Wavelengths in \AA\, of the three main X-ray lines of \ion{C}{v}, \ion{N}{vi}, \ion{O}{vii}, 
\ion{Ne}{ix}, \ion{Mg}{xi} and \ion{Si}{xiii} (from Vainshtein \& Safronova 1978).}
\begin{center}
\begin{tabular}{c@{\ }c@{\ }c@{\ }c@{\ }c@{\ }c@{\ }c@{\ }c@{\ }c@{\ }}
\hline
\hline
{\small label}      & {\small line}                    & {\small \ion{C}{v}}&{\small \ion{N}{vi}}&{\small \ion{O}{vii}}  &{\small \ion{Ne}{ix}} &{\small \ion{Mg}{xi}}  &{\small \ion{Si}{xiii}}\\
\hline
$r$ ({\small $w$})& {\small {\it resonance} }          &{\small 40.279} &{\small 28.792} &{\small 21.603}  &{\small 13.447} &{\small 9.1681} &{\small 6.6471} \\
$i$ ({\small $x$})&{\small {\it  inter-}}        &{\small 40.711} &{\small 29.074} &{\small 21.796}  &{\small 13.548} &{\small 9.2267} &{\small 6.6838} \\
 ~~({\small $y$})     &{\small {\it combination}}       &{\small 40.714} &{\small 29.076} &{\small 21.799}  &{\small 13.551} &{\small 9.2298} &{\small 6.6869} \\
$f$ ({\small  $z$})&{\small {\it  forbidden}}   &{\small 41.464}  &{\small 29.531}  &{\small 22.095}  &{\small 13.697} &{\small 9.3134} &{\small 6.7394} \\
\hline
\hline
\end{tabular}
\end{center}
\end{table}

\begin{figure}
\includegraphics[width=9cm]{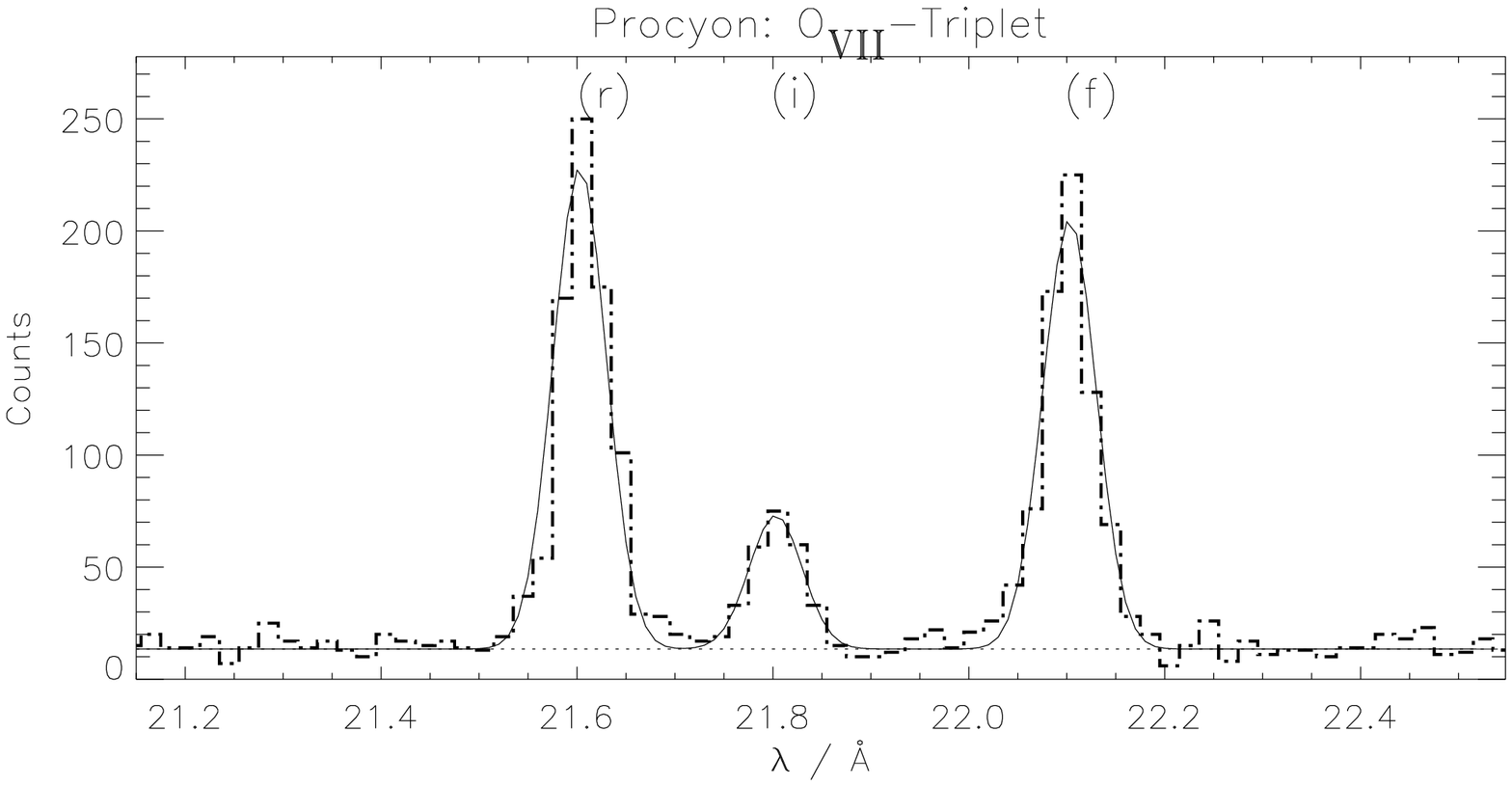} \\
\includegraphics[width=9cm]{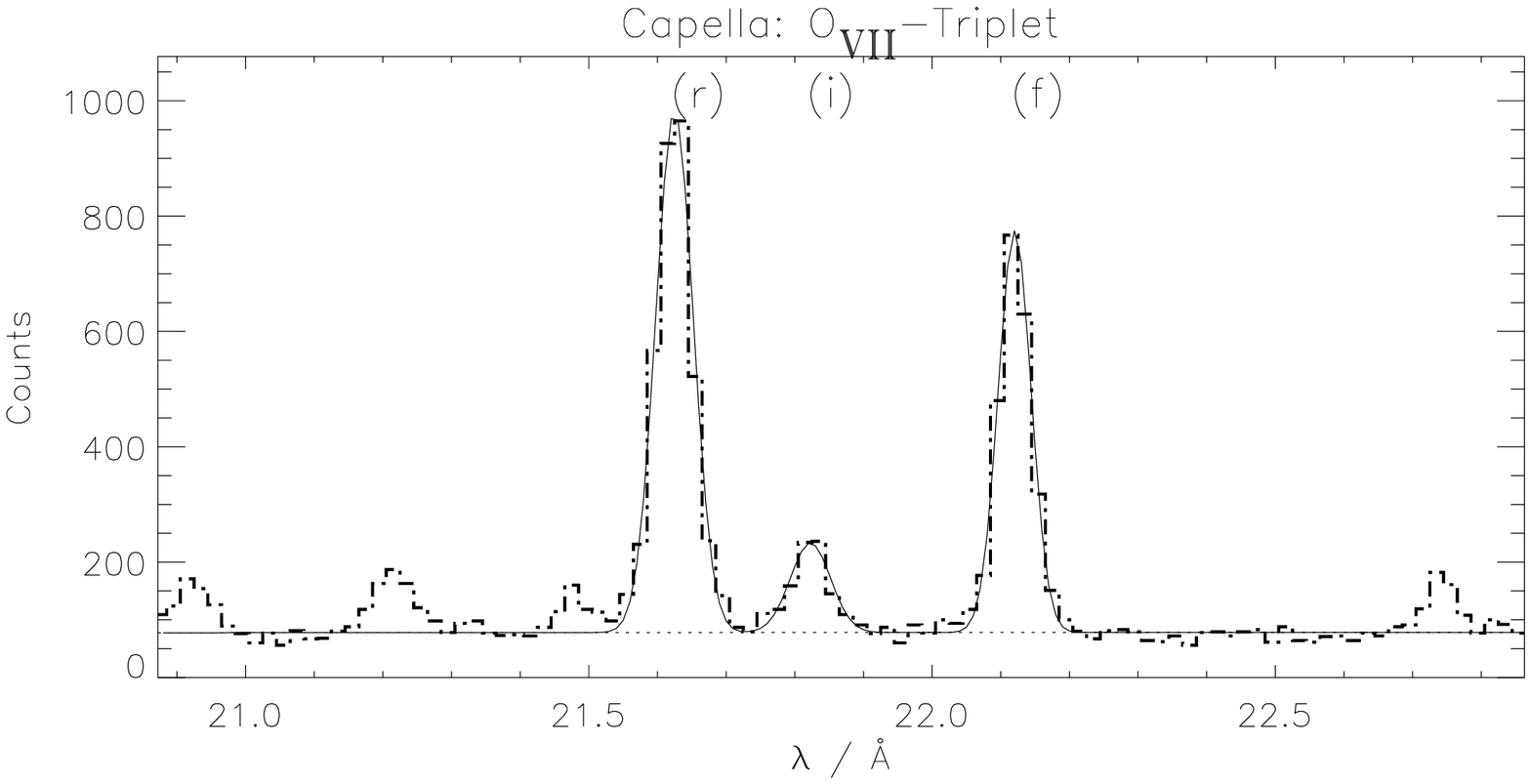} 
\caption{Observed spectra (Chandra, LETGS) of the He-like \ion{O}{vii} triplet 
lines ($r$, $i$, and $f$) of the stellar coronae of 
Capella and Procyon (Ness et al. 2001).}
\label{fig:Capella-Procyon}
\end{figure}

These diagnostics have been recently used
 in numerous observations of extra-solar objects 
thanks to Chandra and XMM-Newton 
 such as  
Active Galactic Nuclei 
(e.g., NGC 5548: Kaastra et al. \cite{jdubau-WB2:Kaastra2000};
NGC 4051: Collinge et al. \cite{jdubau-WB2:Collinge2001};
MCG-6-30-15: Lee et al. \cite{jdubau-WB2:Lee2001};
NGC 4151: Ogle et al \cite{jdubau-WB2:Ogle2000},
Mrk 3: Sako et al. \cite{jdubau-WB2:Sako2000};
NGC 3783: Kaspi et al. \cite{jdubau-WB2:Kaspi2000};
Mkn 509: Pounds et al. \cite{jdubau-WB2:Pounds2001}); 
  stellar coronae and stellar winds
 (e.g., Vela X-1: Schulz et al. \cite{jdubau-WB2:Schulz2002};
 II Pegasi: Huenemoerder et al. \cite{jdubau-WB2:Huenemoerder2001};
Capella: Ness et al. \cite{jdubau-WB2:Ness2001}, Audard et al. \cite{jdubau-WB2:Audard01});
 and X-ray binaries
 (SS 433: Marshall et al. \cite{jdubau-WB2:Marshall2002};
4U 1626-67: Schulz et al.  \cite{jdubau-WB2:Schulz2001};
 EXO 0748-67: Cottam et al. \cite{jdubau-WB2:Cottam2001}). 
As an illustration, the spectra of He-like \ion{O}{vii} ion 
of the stellar coronae of Capella and Procyon are shown in Figure~\ref{fig:Capella-Procyon}.

They allow the determination of the ionization process 
(linked to the temperature)
 in case of the ``Warm Absorber'' in AGNs (photo-ionization and/or an additional 
ionization process such as shock or starburst).
However, these diagnostics should be used cautiously in order to
avoid any mistaken interpretation of the physical parameters of 
the observed plasmas.
Indeed some neglected atomic processes can lead to wrong determination
of the density and of the temperature and of the ionization processes:
 blended dielectronic satellite lines, photo-excitation due to a strong
 UV radiation field, resonant scattering (optical depth). 

\section{Plasma diagnostics}

\begin{figure}
\includegraphics[width=8.9cm,angle=-90]{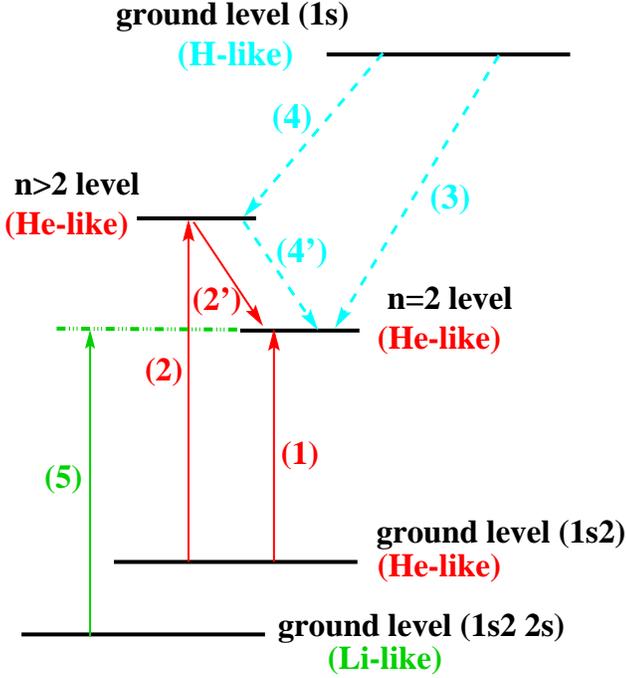}
\caption{Simplified Grotrian diagram reporting the different contributions for the population of
                             a given n=2 shell level. 
                             (1): direct contribution due to collisional excitation (CE) from the ground
                             level (1s$^{2}$) of He-like ions; 
                             (2)+(2'): CE upper level radiative cascade contribution; 
                             (3): direct Radiative Recombination (RR)
                             from H-like ions contribution; 
                             (4)+(4'): RR or Dielectronic Recombination (DR) upper level radiative cascade
                             contribution;
                             and (5): collisional ionization (CI) 
                             from the ground level (1s$^{2}$\,2s) of Li-like ions.
                             Note: CE, CI and DR are only important at high temperature.
                             On the contrary, RR is important at low temperature.}
\label{fig:gotrian2}
\end{figure}

\subsection{Density diagnostic}\label{sec:density}

\indent In the low-density limit, all $n$=2 states are populated directly 
or via upper-level radiative cascades by electron impact from the He-like 
ground state, by recombination (radiative and dielectronic) 
of H-like ions, 
 and for 1s\,2s $^{1}$S$_{0}$ and $^{3}$S$_{1}$ levels, 
by collisional electron ionization from the 1s$^{2}$2s\,$^{2}$S$_{1/2}$ 
lithium-like ground level (see Figure~\ref{fig:gotrian2}). 
This later is only important 
in the out of equilibrium ionization phase such as seen in solar flares 
(e.g., Mewe \& Schrijver \cite{jdubau-WB2:Mewe78a}, 
\cite{jdubau-WB2:Mewe78b}, \cite{jdubau-WB2:Mewe78c}; 
Bely-Dubau \cite{jdubau-WB2:Bely-Dubau82}). 
If the suffix $\lambda$ represents $f$, $i$ and $r$, the line emissivity $\epsilon_{\lambda}$ can
be written as:
$$\epsilon_{\lambda}=n_{e} \left[{\cal C}^{He}_{\lambda}(T_{e})~N(He-like)\right.\hskip3cm $$ 
$$\left. ~~~~+~{\cal C}^{Li}_{\lambda}(T_{e})~N(Li-like) 
~+~\alpha^{H}_{\lambda}(T_{e})~N(H-like)\right] 
$$ 
Where ${\cal C}^{He}_{\lambda}$, ${\cal C}^{Li}_{\lambda}$, and $\alpha^{H}_{\lambda}$
are {\it effective} rate coefficients, i.e including radiative cascades 
inside He-like levels,
respectively excitation, ionization and recombination rates; 
and  $N(He-like)$, $N(Li-like)$, and $N(H-like)$ are
 the population of 1s$^{2}$, 1s$^{2}$2s and 1s levels. 

\begin{figure}[h]
\includegraphics[width=7cm,angle=-90]{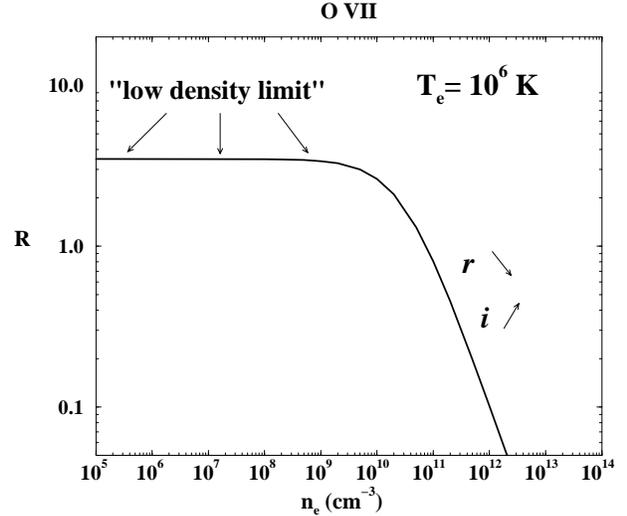}
\caption{R(n$_{e}$)=$f$/$i$ for the He-like ion \ion{O}{vii} in case 
of a collisional plasma. such as stellar coronae 
(calculations are taken from Porquet et al. 2001).}
\label{fig:Rne}
\end{figure}

These $n$=2 levels decay radiatively directly or by cascades to the ground level. 
The relative intensities of the three intense lines are then independent of density.
 As $n_{\mathrm{e}}$ increases from the low-density limit, some of these states 
(1s2s\,$^{3}$S$_{1}$ and $^{1}$S$_{0}$) are depleted by collisions to the nearby states 
where $n_{\mathrm{crit}}$\,C\,=\,A, with C being the sum of the collisional rates depopulating the level, 
A being the radiative transition probability from $n$=2 to $n$=1 (ground state), 
and $n_{\mathrm{crit}}$ being the critical density. Collisional excitation depopulates 
first the 1s2s\,\element[][3]{S}$_{1}$ level (upper level of the {\it forbidden} line) 
to the 1s2p\,\element[][3]{P}$_{0,1,2}$ levels (upper levels of the {\it intercombination} lines). 
The intensity of the {\it forbidden} line decreases while those of the {\it intercombination} 
lines increase, hence implying a reduction of the ratio $R$ (according to Eq.~\ref{eq:RG}), 
over approximately two or three decades of density (see Fig.~\ref{fig:Rne}).

For much higher densities, 1s2s\,$^{1}$S$_{0}$ is also depopulated to 1s2p$^{1}$P$_{1}$, and the 
{\it resonance} line becomes sensitive to the density.\\
 However caution should be taken for low-Z ions 
(i.e. \ion{C}{v}, \ion{N}{vi}, \ion{O}{vii}) since in case of  an intense UV radiation field, 
the photo-excitation between the $^{3}$S term and the $^{3}$P term is not  negligible. 
This process has the same effect on the {\it forbidden} line and on the 
{\it intercombination}  line as the collisional coupling, i.e. lowering of the 
ratio $R$, and thus could mimic a high-density plasma. 
 It should be taken into account to avoid any misunderstanding 
between a high-density plasma and a high  radiation field 
(see e.g. Porquet et al. \cite{jdubau-WB2:Porquet01} for more details).

\subsection{Temperature/ionization process diagnostics}\label{sec:temperature}

\begin{figure}
\includegraphics[width=7cm,angle=-90]{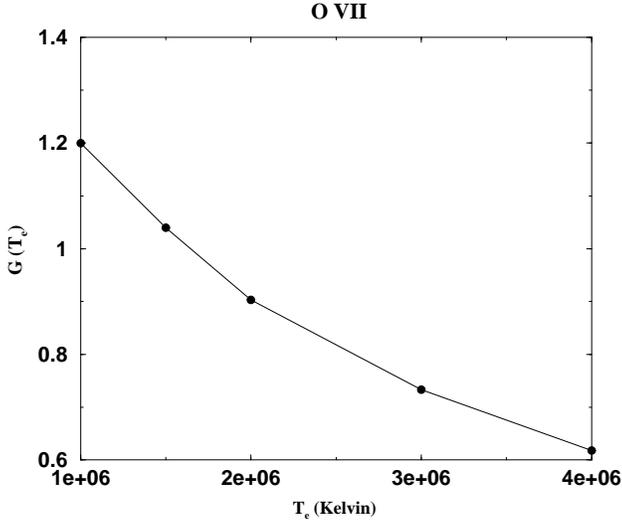} \\
\caption{G(T$_{e}$) for the He-like \ion{O}{vii} in case of a 
collisional plasma, in the low density limit 
(calculations are taken from Porquet et al. 2001).}
\label{fig:G_Te}
\end{figure}

\indent The ratio $G$ (see Eq.~\ref{eq:RG}, and Fig.~\ref{fig:G_Te}) is sensitive to the electron temperature 
since the collisional excitation rates have not the same dependence with  temperature 
for the {\it resonance} line as for the  {\it forbidden} and {\it intercombination} lines.\\
 In addition, as detailed in Porquet \& Dubau \cite{jdubau-WB2:Porquet00} 
(see also Mewe \cite{jdubau-WB2:Mewe99}, and Liedahl \cite{jdubau-WB2:Liedahl99}), 
the relative intensity of the {\it resonance} $r$ line, compared to the 
{\it forbidden} $f$ and the {\it intercombination } $i$ lines, 
contains  information about the ionization processes that occur: a strong  
{\it resonance} line compared to the  {\it forbidden} or the {\it intercombination} 
lines corresponds to collision-dominated plasmas. 
It leads to a ratio of $G=(f+i)/r\sim$1. 
On the contrary, a weak {\it resonance} line corresponds to plasmas dominated  
by photo-ionization ($G=(f+i)/r>$4). \\
However, as mentioned for the density diagnostic, caution should be taken since 
photo-excitation can mimic a hybrid plasmas, i.e. photo-ionization plus collisional ionization, 
e.g. shock or starburst (see $\S$\,4).

\section{Numerical modeling}

\subsection{Atomic data}

Wavelengths and radiative probabilities: since 1930 many calculations 
using different approximations (simple methods up to relativistic many body 
methods). With the observed resolution of Chandra and XMM-Newton, a good agreement
between these calculations can be reached easily, for 
example, for electron excitation rates see Dubau (\cite{jdubau-WB2:Dubau94}).

\subsection{Blended dielectronic satellite lines}
\label{fauthor-E1_sec:fig}

The influence of the blending of dielectronic satellite lines 
for the {\it resonance}, the {\it intercombination} and the 
{\it forbidden} lines has been taken into account where their contribution is not negligible in the 
calculation of $R$ and $G$, affecting the inferred electron temperature and density. 
This is the case  for the high-Z ions produced in a collisional plasma, 
i.e. \ion{Ne}{ix}, \ion{Mg}{xi}, and \ion{Si}{xiii} (Z=10, 12, and 14, respectively).

\begin{equation}
R=\frac{f+satf}{i+sati}
\end{equation}
\begin{equation}
G=\frac{(f+satf)+(i+sati)}{(r+satr)},
\end{equation}
where $satf$, $sati$ and $satr$ are respectively the contribution 
of blended dielectronic satellite lines to the 
{\it forbidden} line, to the {\it intercombination} lines, and to the {\it resonance} line, respectively. 
One can note that at very high density the $^{3}$P levels are depleted to the $^{1}$P level, and in that 
case $i$ decreases and $R$ tends to $satf$/$sati$.\\
The intensity of a dielectronic satellite line arising from a doubly excited state with principal quantum number $n$ 
in a Lithium-like ion produced by dielectronic recombination of a He-like ion  is given by: 
\begin{equation}
I_{s}=N_{\rm He}~ n_{\mathrm e}~ C_{s},
\end{equation}
where $N_{\rm He}$ is the population density of the considered He-like ion in the ground state 1s$^{2}$ with statistical
weight $g_1$ (for He-like ions $g_1=1$).\\
 The rate coefficient (in cm$^{3}$\,s$^{-1}$) for dielectronic recombination is given by (Bely-Dubau et al. 
 \cite{jdubau-WB2:Bely79}):
\begin{equation}
C_{s}=2.0706\ 10^{-16}~\frac{e^{-E_{s}/kT_{\mathrm e}}}{g_{1} T_{\mathrm e}^{3/2}}~F_{2}(s),
\end{equation}
where $E_{s}$ is the energy of the upper level of the satellite 
line $s$ with statistical weight $g_s$ above the  
ground state of the He-like ion.
 $T_{\mathrm e}$ is the electron temperature in K, and 
$F_{2}(s)$ is the so-called line strength factor 
(often of the order of about 10$^{13}$~s$^{-1}$ for the stronger lines) given by
\begin{equation}
F_{2}(s) = {{g_s A_a A_r} \over {(\sum A_a + \sum A_r)}},
\end{equation}
where $A_a$ and
$A_r$ are transition probabilities (s$^{-1}$) by autoionization and radiation,
and the summation is over all possible radiative and auto-ionization 
transitions from the satellite level $s$. \\

At the temperature at which the ion fraction is maximum for the He-like ion 
(see e.g. Mazzotta et al. \cite{jdubau-WB2:Mazzotta98}), the differences between the 
calculations for $R$ (for $G$) with or without taking into account the blended dielectronic satellite lines are only
of about 1$\%$ (9$\%$), $2\%$ (5$\%$), and 5$\%$ (3$\%$) for \ion{Ne}{ix}, \ion{Mg}{xi}, and 
\ion{Si}{xiii} at the low-density limit and for $T_{\mathrm rad}$=0\,K, respectively
(see Porquet et al. \cite{jdubau-WB2:Porquet01}). 
At lower temperature, the contribution of the blended satellite lines 
become larger and at higher temperature it can be neglected.\\

For photo-ionized plasmas where recombination prevails 
and the temperature is much lower (e.g., T$\la$0.1T$_{m}$), 
the effect on $R$ and $G$ can be much bigger since 
I$_{sat}$/I$_{r}$$\propto T^{-1}{\rm e}^{{(E_{r}-E_{sat}})/k{\rm T}}$.
For very high density $n_{\mathrm e}$ the contribution of the blended 
dielectronic satellite lines to the forbidden line leads to a ratio $R$ which 
tends to $satf/sati$, hence decreases much slower with $n_{\mathrm e}$ 
than in the case where the contribution of the blended DR satellites is not taken into account.
The importance of the dielectronic satellite lines can be seen in the 
\ion{O}{vii} He-like triplet spectra of Procyon on Figure~2, 
on the right hand side of $r$ and the left hand side of $f$. 
They indicate a lower temperature coronal plasma compared to the Capella spectra. 

\subsection{Optical depth}

If the optical depth of the resonance line is not taken into account, 
the calculated ratio G could be overestimated (inferred temperature underestimated) 
when the optically-thin approximation is no longer valid. 
This has been estimated with an {\it escape-factor method}, e.g., 
for the case of a {\it Warm Absorber in an AGNs} 
(Porquet, Kaastra, Mewe, Dubau \cite{jdubau-WB2:Porquet02}).

\section{Photo-ionized model}

\begin{figure}[t]
\includegraphics[width=6.75cm,angle=90]{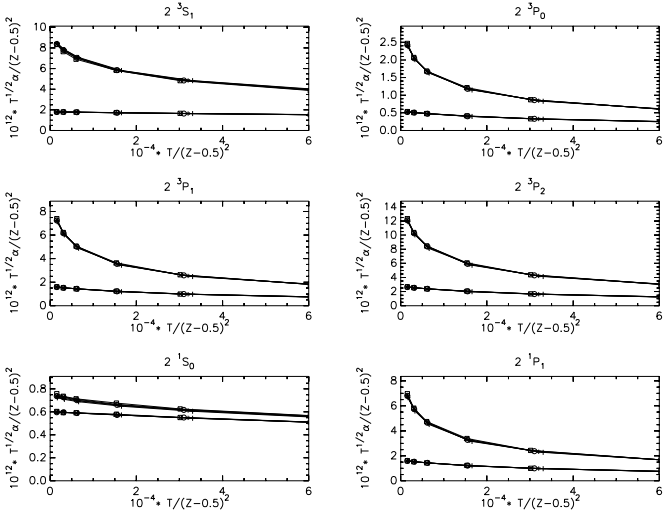} 
\caption{Scaled total radiative recombination rates 
(upper curves: direct plus cascade contribution from n$>$2 levels) 
$\alpha^{\mathrm{s}}$=T$^{1/2}$\,$\alpha$/($Z$-0.5)$^{2}$ 
($\times$10$^{12}$ cm$^{3}$\,s$^{-1}$) versus 
T$^{\mathrm{s}}$=T/($Z$-0.5)$^{2}$ ($\times$10$^{-4}$) 
towards each $n$=2 level (Plus, star, circle and cross are respectively 
for $Z$=8,10,12,14), and for comparison the direct contribution 
(lower curve in each graph). T is in Kelvin.}
\label{fig:rr}
\end{figure}

\begin{figure}[t]
\includegraphics[width=4.5cm,angle=90]{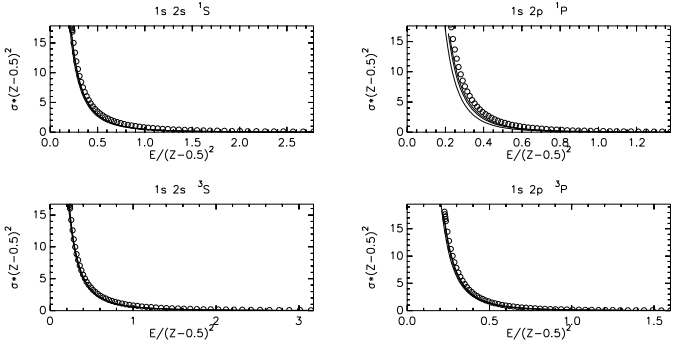} 
\caption{Scaled photoionization cross sections 
$\sigma_{\mathrm{s}}$=$\sigma$\,($Z$-0.5)$^{2}$ (in cm$^{-3}$\,s$^{-1}$) 
as a function of E/($Z$-0.5)$^{2}$ (E is in Rydberg). {\it Empty circles}: 
photo-ionization cross sections calculated in 
Porquet \& Dubau (2000); {\it solid lines}: 
photo-ionization cross sections available in Topbase for different
 values of $Z$=6, 10, 14.}
\label{fig:cross-sections}
\end{figure}

``Warm Absorbers'' in AGN are typical examples of photo-ionized (PI) plasmas
 which are dominantly ionized by an intense radiative source, in such plasmas
 the electron temperature is relatively small. 
Then atomic levels are predominantly populated by
 direct (radiative and/or dielectronic) recombination or by radiative cascades from upper levels,
which are very important data.
For the modeling of such plasmas the atomic data required are the radiative probabilities,
the radiative and dielectronic recombination rates to individual levels, 1s\,$n$l\,$^{S}$L$_{J}$,
with n$\leq$10, plus extrapolation of $n$ to infinity.
Such a large basis of ionic states is necessary because the radiative recombination is
slowly convergent with $n$.\\
The cascade contribution of radiative contribution is very important for small temperature,
as it can be seen in Figure~\ref{fig:rr}. 
In Figure~\ref{fig:cross-sections} the comparison of two different calculations 
for radiative recombination is done:\\
(1) TOPBASE, a very sophisticated calculation using the R-matrix code;\\
(2) screened hydrogenic data obtained from analytical quantum 
expression. It is interesting to see in Figures~\ref{fig:rr} 
and \ref{fig:cross-sections} that the data can be easily scaled
along the iso-electronic sequence 
(Z=8, 10, 12, 14: Porquet \& Dubau \cite{jdubau-WB2:Porquet00}). 

\begin{figure}[t]
\includegraphics[width=6.5cm,angle=-90]{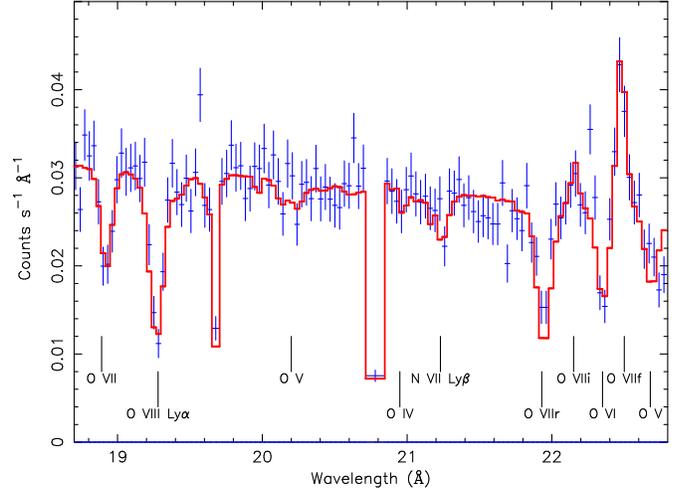}
\caption{X-ray spectra (XMM-Newton, RGS) of the energy range containing the He-like \ion{O}{vii} 
triplet ($r$, $i$, and $f$ lines) showing  also the photo-absorbed lines 
of Ly-$\alpha$ (\ion{O}{viii}), $r$ (\ion{O}{vii}) as well two satellite lines 
$q$ (\ion{O}{vi}) and $\beta$ (\ion{O}{v}) (Steenbrugge et al. 2002).}
\label{fig:NGC5548}
\end{figure}

As detailed above the electron density diagnostic is due
to the electron excitation inside the n=2 levels. 
However to get reliable diagnostics, photo-excitation 
between these close-levels must also be accounted for.

Recently,  Kinkhabwala et al.  (\cite{jdubau-WB2:Kinkhabwala02}), 
 pointed out also the important effect of
the photo-excitation in the high Rydberg series lines.  Indeed, the ratio of high-$n$ 
lines to Ly$\alpha$ bring evidence for photo-excitation in Warm Absorber Seyfert\,2. 
They clearly showed that in addition to photo-ionization 
(treated in Porquet \& Dubau \cite{jdubau-WB2:Porquet00}), 
 photo-excitation process is sufficient to fit the data of Seyfert  galaxies 
without needing an additional collisional ionization process (e.g. shock or starburst).
 Then both photo-ionization and photo-excitation are needed to inferred 
unambiguously the ionization process occurring in the plasmas.

\begin{figure}[t]
\hspace*{0.05cm}\includegraphics[width=11.cm,angle=-90]{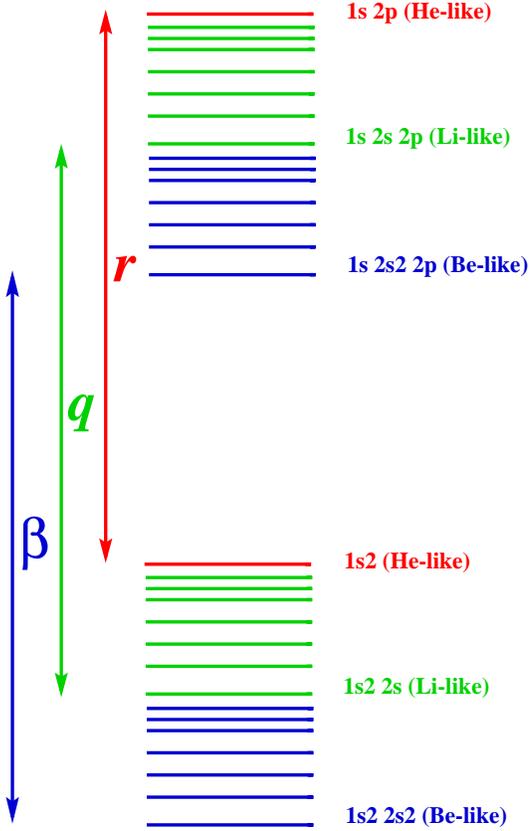}
\caption{Energy level diagram showing the parent resonance line $r$ in the
Helium-like ion together with two strong satellite lines due
to photo-absorption: $q$ in Li-like, $\beta$ in Be-like.}
\label{fig:DR}
\end{figure}
 
In Figure~\ref{fig:NGC5548} is shown an interesting  photo-ionized plasma 
spectrum of \ion{O}{vii} and \ion{O}{viii} obtained by XMM-Newton RGS 
for NGC 5548 (Steenbrugge et al. \cite{jdubau-WB2:Steenbrugge2002}). 
On this spectra one can see photo-absorbed lines 
of Ly-$\alpha$ (\ion{O}{viii}), $r$ (\ion{O}{vii}) as well two 
satellite lines $q$ (\ion{O}{vi}) and $\beta$ (\ion{O}{v}).
On the contrary, the $f$ and $i$ lines are seen as expected in emission,
they are probably due to recombination of H-like ion (here \ion{O}{viii}). 
The two  satellite lines $q$ and $\beta$ 
 are located respectively between between \ion{O}{vii}
 $i$ and $f$, and on the right hand side of $f$.
An energy level diagram is shown in Figure~\ref{fig:DR}, 
it explains why the $q$ and $\beta$ lines are satellites 
of $r$, i.e. have very close in wavelength to $r$.
We can also notice that the upper levels of $q$ and $\beta$
can auto-ionize since they are above 1s$^{2}$: 
1s\,2s\,2p decays to 1s$^{2}$ 
and 1s\,2s$^{2}$\,2p decays preferably to 1s$^{2}$\,2s and 1s$^{2}$\,2p.

\section{Collisional and hybrid models}

In pure collisional model, 
atoms are ionized by electron collision, 
the free electrons being heated by some external source.
Electron collision is the dominant excitation process,
and spontaneous emission is the dominant de-excitation process. 

A hybrid model is a mixing of both collisional and photo-ionized models.

\noindent A strong radiation field can mimic a high density if the 
photo-excitation $^{3}$S$_1$ level ({\it f} line) $\to$ $^{3}$P$_{0,1,2}$ levels 
({\it i} lines) exceeds the electron collisional excitation. ex: $\zeta$ Puppis 
(Kahn et al. \cite{jdubau-WB2:Kahn01}, Cassinelli et al. 
\cite{jdubau-WB2:Cassinelli01}). Rate of photo-excitation (in s$^{-1}$) 
(Mewe \& Schrijver \cite{jdubau-WB2:Mewe78a}) in a stellar photospheric 
radiation field with effective black-body radiation temperature ${T_{\mathrm rad}}$ is written
 as: 
\begin{equation}
B_{mp_k} = {\frac{W A_{p_km} (w_{p_k}/w_m)}{exp \Bigl(\frac{\Delta E_{mp_k}}{kT_{\mathrm rad}}\Bigr) - 1}},
\end{equation}

\noindent where $A$ and $B$ are the Einstein coefficients and the radiation is diluted by a factor $W$ given 
by
\begin{equation}
W=\frac{1}{2}~\left[1-\left(1-\left(\frac{r_{*}}{r}\right)\right)^{1/2}\right],
\end{equation}

\noindent $\bullet$ {\bf W=1/2} (close to the stellar surface, $r=r_*$; e.g., 
\object{Capella} and \object{Procyon}: Audard et al. \cite{jdubau-WB2:Audard01}, 
Mewe et al. \cite{jdubau-WB2:Mewe01}, Ness et al. \cite{jdubau-WB2:Ness2001}).\\
$\bullet$ {\bf W$<<$ 1/2} (radiation originates from another star at larger distance; 
e.g., Algol, where K-star is irradiated by B-star, $W \simeq 0.01$: Ness et al. \cite{jdubau-WB2:Ness02}).\\ 
%
Porquet et al. (\cite{jdubau-WB2:Porquet01}) showed that photo-excitation 
is important for \ion{C}{v}, \ion{N}{vi}, \ion{O}{vii} 
for T$_{\mathrm rad}\geq$(5-10) 10$^3$\,K 
(see Fig.~3 in Porquet et al. \cite{jdubau-WB2:Porquet02}), 
and for higher-Z ions when T$_{\mathrm rad}\geq$ few 10$^4$~K.\\

 \section{Laboratory spectra}

Helium-like lines are now currently observed in many laboratory 
plasmas: tokamaks (e.g., Doyle \& Schwob \cite{jdubau-WB2:Doyle82}), 
laser produced plasmas (e.g., Renaudin et al. \cite{jdubau-WB2:Renaudin94}), 
Z-pinch (e.g., Coulter et al. \cite{jdubau-WB2:Coulter88}).
The former diagnostics have often been used.

Atomic data can be checked by comparison with
laboratory measurements, ions being produced by ionic 
sources: EBIT (Wargelin et al. \cite{jdubau-WB2:Wargelin2001}).

\section{Conclusion}

We have shown that the ratios of the three main lines 
(forbidden, intercombination and resonance) 
of He-like ions provide very powerful diagnostics 
for totally or partially photo-ionized media. 
For the first time, these diagnostics can be applied to 
non-solar plasmas thanks to the high spectral resolution 
and the high sensitivity of the new
 X-ray satellites {\sl Chandra}, and {\sl XMM-Newton}:\\
$\bullet$ collisional plasmas: 
{\it e.g., stellar coronae (OB stars, late type stars, active stars, ...)}; \\
$\bullet$  photo-ionized or hybrid plasmas (photo-ionization + collisional ionization): \\
{\it e.g., ``Warm Absorber'' (in AGNs), X-ray binaries, ...}\\
$\bullet$ out of equilibrium plasmas: {\it e.g., SNRs, stellar flares, ...}.\\
 
These diagnostics have strong advantages.
The lines are emitted by the same ionization stage of one element, 
thus eliminating any uncertainties due to elemental abundances. 
In addition, since the line energies are relatively close together, this minimizes
wavelength dependent instrumental calibration uncertainties, 
thus ensuring that observed photon count rates can be used almost directly. 

These diagnostics should be used not as stand-alone one but
should be used combined to other plasma diagnostics such as those
based on radiative recombination
continuum (RRC; Liedahl \& Paerels \cite{jdubau-WB2:Liedahl96}), 
Fe-L shell lines (see review of Liedahl et al. \cite{jdubau-WB2:Liedahl92}), 
which give respectively indication
on the ionization process as well as on the density.\\
 In a close future high spectral resolution spectra at higher energy range 
 (Astro-E2, Constellation-X, XEUS) will give access to higher Z He-like ions
such as sulfur, calcium and iron, which are sensitive to higher
range of density and temperature. 

\begin{acknowledgements}

D.P. acknowledges grant support from the ``Institut National des Sciences de l'Univers'' and from the
``Centre National \linebreak d'Etudes Spatial'' (France). 
\end{acknowledgements}



\begin{thebibliography}{}


\bibitem[2001]{jdubau-WB2:Audard01}
Audard, M., Behar, E., G{\"u}del, M., Raassen, A. J. J.,
 Porquet, D., Mewe, R., Foley, C. R., Bromage, G. E. (2001), A\&A, 365, L329
\bibitem[1979]{jdubau-WB2:Bely79}
Bely-Dubau, F., Gabriel, A. H., Volont{\'e}, S. (1979), MNRAS, 189, 801
\bibitem[1982]{jdubau-WB2:Bely-Dubau82} 
Bely-Dubau F., Dubau J., Faucher P., Gabriel A.~H. (1982), MNRAS,  198, 239
\bibitem[1981]{jdubau-WB2:Burek81} 
Burek A.~J., Barrus D.~M., Blake R.~L., Fenimore E.~E. (1981), ApJ,  243, 660
\bibitem[2001]{jdubau-WB2:Cassinelli01}
 Cassinelli J. P., Miller N. A., Waldron W. L., 
MacFarlane J. J., Cohen D. H. (2001), ApJ, 554, L55
\bibitem[2001]{jdubau-WB2:Collinge2001} 
Collinge M.~J., Brandt W.~N., Kaspi S., Crenshaw, D. M., 
Elvis, M., Kraemer, S. B., Reynolds, C. S., Sambruna, R. M.,
 Wills, Beverley J. (2001), ApJ,  557, 2
\bibitem[2001]{jdubau-WB2:Cottam2001} 
Cottam J., Kahn S.~M., Brinkman A.~C., 
den Herder J.~W., Erd C. (2001), A\&A,  365, L277
\bibitem[1988]{jdubau-WB2:Coulter88} 
Coulter M.~C., Apruzese J.~P., Kepple P.~C. (1988), 
Journal of Applied Physics,  63, 2221
\bibitem[1981]{jdubau-WB2:Culhane81} 
Culhane J.~L., Rapley C.~G., Bentley R.~D., Gabriel, A. H.,
 Phillips, K. J., Acton, L. W., Wolfson, C. J., Catura, R. C.,
 Jordan, C., Antonucci, E. (1981), ApJ,  244, L141
\bibitem[1980]{jdubau-WB2:Doyle80}
Doyle J.G. (1980), A\&A, 87, 183
\bibitem[1982]{jdubau-WB2:Doyle82} 
Doyle J.~G., Schwob J.~L., 1982, Journal of Physics B Atomic Molecular Physics,  15, 813
\bibitem[1972]{jdubau-WB2:Doschek72} 
Doschek G.~A., Meekins J.~F., Cowan R.~D. (1972), ApJ,  177, 261
\bibitem[1979]{jdubau-WB2:Doschek79} 
Doschek G.~A., Kreplin R.~W., Feldman U. (1979), ApJ,  233, L157
\bibitem[1994]{jdubau-WB2:Dubau94} 
Dubau J. (1994), Atomic Data and Nuclear Data Tables,  57, 21
\bibitem[1969]{jdubau-WB2:Gabriel69}
Gabriel A.H. \& Jordan, C. (1969), MNRAS, 145, 241
\bibitem[1973]{jdubau-WB2:Grineva73} 
Grineva Y.~I., Karev V.~I.~K.~V.~V., Krutov V.~V., Mandelstam S.~L., 
Vainstein L.~A., Vasilyev B.~N., Zhitnik I.~A. (1973), Solar Physics,  29, 441
\bibitem[1996]{jdubau-WB2:Harra-Murnion96} 
Harra-Murnion, L.~K., Phillips, K. J. H., Lemen, J. R., 
Zarro, D. M., Greer, C. J., Foster, V. J., Barnsley, R., 
Coffey, I. H., Dubau, J., Keenan, F. P., Fludra, A., 
Rachlew-Kaellne, E., Watanabe, T., Wilson, M. (1996), A\&A 308, 670-684
\bibitem[2001]{jdubau-WB2:Huenemoerder2001} 
Huenemoerder D.~P., Canizares C.~R., Schulz N.~S. (2001), ApJ,  559, 1135
\bibitem[2000]{jdubau-WB2:Kaastra2000} 
Kaastra J.~S., Mewe R., Liedahl D.~A., Komossa S., 
Brinkman A.~C. (2000), A\&A,  354, L83
\bibitem[2002]{jdubau-WB2:Kaastra02}
Kaastra J. S, Mewe R., Porquet D., Raassen A. J. J. (2002), 
in preparation (Paper IV)
\bibitem[2001]{jdubau-WB2:Kahn01}
Kahn S.~M., Leutenegger, M. A., Cottam, J., Rauw, G.,
 Vreux, J.-M., den Boggende, A. J. F., 
Mewe, R., G{\"u}del, M. (2001), A\&A, 365, L312
\bibitem[2000]{jdubau-WB2:Kaspi2000} 
Kaspi S., Brandt W.~N., Netzer H., Sambruna R., 
Chartas G., Garmire G.~P., Nousek J.~A. (2000), ApJ,  535, L17
\bibitem[1997]{jdubau-WB2:Kato97} 
Kato T., Safronova U., Shlypteseva A., 
Cornille M., Dubau J., Nilsen J., 1997, 
Atomic Data and Nuclear Data Tables,  67, 225
\bibitem[2002]{jdubau-WB2:Kinkhabwala02}
Kinkhabwala et al. (2002), these proceedings
\bibitem[2001]{jdubau-WB2:Lee2001} 
Lee J.~C., Ogle P.~M., Canizares C.~R., Marshall H.~L., Schulz N.~S., 
Morales R., Fabian A.~C., Iwasawa K. (2001), ApJ,  554, L13
\bibitem[1992]{jdubau-WB2:Liedahl92} 
Liedahl D.~A., Kahn S.~M., Osterheld A.~L., Goldstein W.~H., 1992, ApJ,  391, 306
\bibitem[1996]{jdubau-WB2:Liedahl96}
Liedahl, D. A.,  Paerels, F. (1996), ApJ, 468, 33
\bibitem[1999]{jdubau-WB2:Liedahl99}
Liedahl D.~A. (1999), X-Ray Spectroscopy in Astrophysics,  189
\bibitem[2002]{jdubau-WB2:Marshall2002} 
Marshall H.~L., Canizares C.~R., Schulz N.~S. (2002), ApJ,  564, 941
\bibitem[1998]{jdubau-WB2:Mazzotta98}
 Mazzotta P., Mazzitelli, G., Colafrancesco, S.,
 Vittorio, N. (1998), A\&AS, 133, 403
\bibitem[2001]{jdubau-WB2:Mauche2001} 
Mauche C.~W., Liedahl D.~A., Fournier K.~B., 2001, ApJ,  560, 992
\bibitem[1980]{jdubau-WB2:McKenzie80} 
McKenzie D.~L., Landecker P.~B., Broussard R.~M., 
Rugge H.~R., Young R.~M., Feldman U., Doschek G.~A. (1980), ApJ,  241, 409
\bibitem[1978a]{jdubau-WB2:Mewe78a}
Mewe R. \& Schrijver J. (1978a), A\&A, 65, 99
\bibitem[1978b]{jdubau-WB2:Mewe78b}
Mewe R. \& Schrijver J. (1978b), A\&A, 65, 115 
\bibitem[1978c]{jdubau-WB2:Mewe78c}
 Mewe R. \& Schrijver J. (1978c), A\&AS, 45, 11 
\bibitem[1999]{jdubau-WB2:Mewe99}
Mewe R. (1999), X-Ray Spectroscopy in Astrophysics,  109
\bibitem[2001]{jdubau-WB2:Mewe01}
 Mewe R., Raassen A.J.J., Drake J.J., Kaastra J.S., 
van der Meer R.L.J., Porquet D. (2001), A\&A, 368, 888 
\bibitem[2002]{jdubau-WB2:Mewe02}
Mewe R. (2002), these proceedings
\bibitem[2001]{jdubau-WB2:Ness2001}
Ness J.-U., Mewe R., Schmitt J.H.M.M., 
Raassen A.J.J., Porquet D., Kaastra, J. S.,
 van der Meer, R. L. J., Burwitz, V., Predehl, P. (2001), A\&A, 367, 282
\bibitem[2002]{jdubau-WB2:Ness02}
Ness J.-U. et al. (2002), in preparation
\bibitem[1973]{jdubau-WB2:Neupert73} 
Neupert W.~M., Swartz M., Kastner S.~O. (1973), Solar physics,  31, 171
\bibitem[2000]{jdubau-WB2:Ogle2000} 
Ogle P.~M., Marshall H.~L., Lee J.~C., Canizares C.~R. (2000), ApJ,  545, L81
\bibitem[1975]{jdubau-WB2:Parkinson75} 
Parkinson J.~H. (1975), Solar physics,  42, 183
\bibitem[1978]{jdubau-WB2:Parkinson78} 
Parkinson J.~H., Wolff R.~S., Kestenbaum H.~L., Ku, W. H.-M.,
 Lemen, J. R., Long, K. S., Novick, R., Suozzo, R. J.,
 Weisskopf, M. C. (1978), Solar physics,  60, 123
\bibitem[1982]{jdubau-WB2:Phillips82} 
Phillips K.~J.~H., Fawcett B.~C., Kent B.~J., Gabriel, A. H.,
 Leibacher, J. W., Wolfson, C. J., Acton, L. W., Parkinson, J. H.,
 Culhane, J. L., Mason, H. E. (1982), ApJ,  256, 774
\bibitem[2000]{jdubau-WB2:Porquet00}
Porquet D. \& Dubau J. (2000), A\&AS, 143, 495 (Paper I)
\bibitem[2001]{jdubau-WB2:Porquet01}
Porquet D., Mewe R., Dubau J., Raassen A. J. J., Kaastra J. S (2001), A\&A, 376, 1113 (Paper II)
\bibitem[2002]{jdubau-WB2:Porquet02}
 Porquet D., Kaastra J. S, Mewe R., Dubau J. (2002), in preparation (Paper III)
\bibitem[2001]{jdubau-WB2:Pounds2001} 
Pounds K., Reeves J., O'Brien P., Page K., 
Turner M., Nayakshin S. (2001), ApJ,  559, 181
\bibitem[1981]{jdubau-WB2:Pradhan81}
Pradhan A. K. \& Shull J. M. (1981), ApJ, 249, 821
\bibitem[1977]{jdubau-WB2:Pye77} 
Pye J.~P., Evans K.~D., Hutcheon R.~J. (1977), MNRAS,  178, 611
\bibitem[1994]{jdubau-WB2:Renaudin94} 
Renaudin P., Chenais-Popovics C., Gauthier J.~C., 
Peyrusse O., Back C.~A. (1994), Physical Review E,  50, 2186
\bibitem[2000]{jdubau-WB2:Sako2000} 
Sako M., Kahn S.~M., Paerels F., Liedahl D.~A. (2000), ApJ,  543, L115
\bibitem[2001]{jdubau-WB2:Schulz2001} 
Schulz N.~S., Chakrabarty D., Marshall H.~L., Canizares C.~R., 
Lee J.~C., Houck J. (2001), ApJ,  563, 941
\bibitem[2002]{jdubau-WB2:Schulz2002} 
Schulz N.~S., Canizares C.~R., Lee J.~C., Sako M. (2002), ApJ,  564, L21
\bibitem[2002]{jdubau-WB2:Steenbrugge2002} 
Steenbrugge, J. S. Kaastra, A. C. Brinkman, R. Edelson (2002), these proceedings
\bibitem[1978]{jdubau-WB2:Vainshtein78}
Vainshtein L.~A., Safronova U.~I. (1978), 
Atomic Data and Nuclear Data Tables,  21, 49
\bibitem[1970]{jdubau-WB2:Walker70} 
Walker A.~B.~C., Rugge H.~R. (1970), A\&A,  5, 4
\bibitem[2001]{jdubau-WB2:Wargelin2001} 
Wargelin B.~J., Kahn S.~M., Beiersdorfer P., 2001, Phys. Rev. A,  63, 2710
\end{thebibliography}
\end{document}